\documentclass{article}
\usepackage{spconf,amsmath,graphicx,hyperref}

\usepackage{comment}

\usepackage{booktabs}
\usepackage{multirow}
\usepackage{array}
\usepackage{arydshln}
\usepackage{hyperref}
\usepackage{xcolor}

\newcommand{\deemph}[1]{{\color{black!40}#1}}

\title{Domain-Invariant Representation Learning of Bird Sounds}
\name{Ilyass Moummad$^{1}$, Romain Serizel$^{2}$, Emmanouil Benetos$^{3}$, Nicolas Farrugia$^{4}$}
\address{
    $^{1}$ INRIA, LIRMM, Université de Montpellier, Montpellier, France \\
    $^{2}$ Université de Lorraine, Loria, Nancy, France \\
    $^{3}$ C4DM, Queen Mary University of London, London, UK \\
    $^{4}$ IMT Atlantique, Lab-STICC, Brest, France \\
}
\begin{document}
\ninept
\maketitle
\begin{abstract}
Passive acoustic monitoring (PAM) is crucial for bioacoustic research, enabling non-invasive species tracking and biodiversity monitoring. Citizen science platforms provide large annotated datasets from focal recordings, where the target species is intentionally recorded. However, PAM requires monitoring in soundscapes, creating a domain shift between focal and passive recordings, challenging deep learning models trained on focal recordings. To address domain generalization, we leverage supervised contrastive learning by enforcing domain invariance across same-class examples from different domains. Additionally, we propose ProtoCLR, an alternative to SupCon loss which reduces the computational complexity by comparing examples to class prototypes instead of pairwise comparisons. We conduct few-shot classification based on BIRB, a large-scale benchmark to assess pre-trained bioacoustic models. Our findings suggest that ProtoCLR is a better alternative to SupCon.\footnote{Code: \url{https://github.com/ilyassmoummad/ProtoCLR}.\\This work was granted access to the HPC resources of IDRIS under the allocation 2024-AD011015097 made by GENCI.}
\end{abstract}
\section{Introduction}
\label{sec:intro}

Passive Acoustic Monitoring (PAM) is a non-invasive method for studying wildlife through sound. By using acoustic recorders, researchers can gather data on animal behavior, migration, and population trends without disturbance~\cite{pam}. PAM is useful for monitoring endangered species, offering long-term insights for conservation.

In recent years, deep learning models have emerged as a powerful tool to process and analyze complex bioacoustic data~\cite{deepbioac}. A key source of training data for these models comes from citizen science platforms like Xeno-Canto (XC)~\cite{xc}, which contains over one million annotated vocalizations from more than 10,000 species, primarily birds. These citizen-led initiatives have significantly expanded the availability of labeled wildlife sound data, enabling the development of more robust and accurate deep learning models~\cite{birdnet}.

However, a challenge arises from the difference between data collected on platforms like XC and PAM recordings. In citizen science platforms, recordings are typically focal, where the recorder is aimed directly at the species of interest. In contrast, PAM systems passively capture sounds within natural soundscapes, leading to recordings containing a mix of species vocalizations and background environmental noise. This difference in recording conditions creates a domain shift, complicating the ability of models trained on focal data to generalize to soundscape recordings in PAM. 

Supervised contrastive learning (SupCon)~\cite{supcon}, a supervised learning framework for training robust feature extractors, has demonstrated strong generalization capabilities for transfer learning in bioacoustics, particularly in few-shot classification~\cite{moummadicasspw} and detection~\cite{rcl}. However, these studies have been limited to settings where both training and testing rely on focal recordings, and therefore do not address the domain shift challenge associated with testing on PAM recordings when models are trained on focal recordings.

Domain Generalization~(DG)~\cite{dg} aims to develop models that learn robust features that are domain-invariant, i.e. capable of generalizing to new unseen domains without prior knowledge or access to target domain data during training. SupCon offers a promising approach for DG. In SupCon, the objective is to learn an embedding space where same-class examples are pulled together and different-class examples are pushed apart. This clustering can promote domain-invariance when sufficient domain diversity is present in the dataset, allowing the model to focus on features that are domain-invariant, without relying on any domain annotations. In contrast, its self-supervised counterpart SimCLR~\cite{simclr} lacks this explicit mechanism for domain-invariance, as it relies solely on augmentations to create positive pairs. Without label information, SimCLR requires carefully designed augmentations that account for domain shift~\cite{udg}. 

Despite its effectiveness, SupCon is computationally expensive due to the need for pairwise similarity calculations between all examples. To address this, in this paper we introduce ProtoCLR (Prototypical Contrastive Learning of Representations), a more efficient variant. By analyzing SupCon's gradient and drawing inspiration from the generalization capabilities of prototypical learning~\cite{protonets, protonce} in few-shot learning, ProtoCLR replaces pairwise comparisons with a prototypical contrastive loss that compares examples to class prototypes (one prototype per class), retaining the original objective while significantly reducing computational complexity.

We propose a new few-shot classification evaluation based on the BIRB~\cite{birb} benchmark to evaluate the generalization capabilities of models trained on XC's focal recordings and tested on diverse soundscape datasets. This benchmark is designed to assess how well models can generalize across domains in challenging few-shot scenarios. 
We validate our proposed loss ProtoCLR on this benchmark, demonstrating its effectiveness in improving DG in bird sound classification.

In this work, we make the following contributions:
\begin{itemize} 
\item We establish a large-scale few-shot benchmark for bird sound classification using BIRB datasets, evaluating model generalization from focal to soundscape recordings.
\item We introduce ProtoCLR, a novel supervised contrastive loss that reduces computational complexity of SupCon by using class prototypes instead of pairwise comparisons.
\end{itemize}

\section{Related Work}

Nolasco et al.~\cite{nolasco5shots} reformulate bioacoustic sound event detection using a few-shot learning approach to recognize species from a few labeled examples, making it suitable for rare species but limited to single-species detection per task. Heggan et al.~\cite{metaaudio} introduce MetaAudio, a few-shot benchmark for audio classification, including BirdCLEF 2020~\cite{birdclef2020}, which focuses on generalizing to new classes but only includes focal recordings.

To address the generalization challenge from focal to soundscape recordings, the BIRB~\cite{birb} benchmark focuses on few-shot retrieval, retrieving labeled sounds from large, unlabeled datasets. BirdSet~\cite{birdset} emphasizes transfer learning, evaluating models across various downstream classification tasks.

DG~\cite{dg} has emerged as a critical approach to tackle domain shift, where the test data distribution differs from the training data. It aims to learn robust, domain-invariant representations using only source domain data.
DG methods typically focus on learning domain-invariant representations, using techniques like domain alignment, meta-learning and data augmentation~\cite{dg}. These approaches help the model learn features that remain consistent across varying domains. In bioacoustics, DG is especially important due to the difficulty in collecting annotated soundscape recordings compared to focal data~\cite{birb, birdset}.

Invariant learning has gained attention, where models are trained to learn features that remain invariant across different variations in data, such as augmented versions in self-supervised learning~\cite{simclr, dino, bt} or same-class examples in supervised learning~\cite{supcon}. Moummad et al.~\cite{moummadicasspw} showed that invariant learning effectively learns representations of bird sounds that perform well in few-shot learning tasks for recognizing new species not seen during training. However, their experiments focus on 5-way 1-shot classification tasks, that do not represent realistic scenarios.
In our study, we examine test datasets featuring passively recorded sounds from a variable number of species, reflecting more realistic and unencountered training scenarios.

\section{Method}

In this section, we revisit the supervised contrastive loss SupCon and derive ProtoCLR, an alternative that integrates class-level prototypes into the supervised contrastive learning paradigm. We discuss the computational advantages of ProtoCLR and show that they optimize for a similar objective near convergence.

\begin{figure}[!h]
\centering
\centerline{\includegraphics[width=0.5\textwidth]{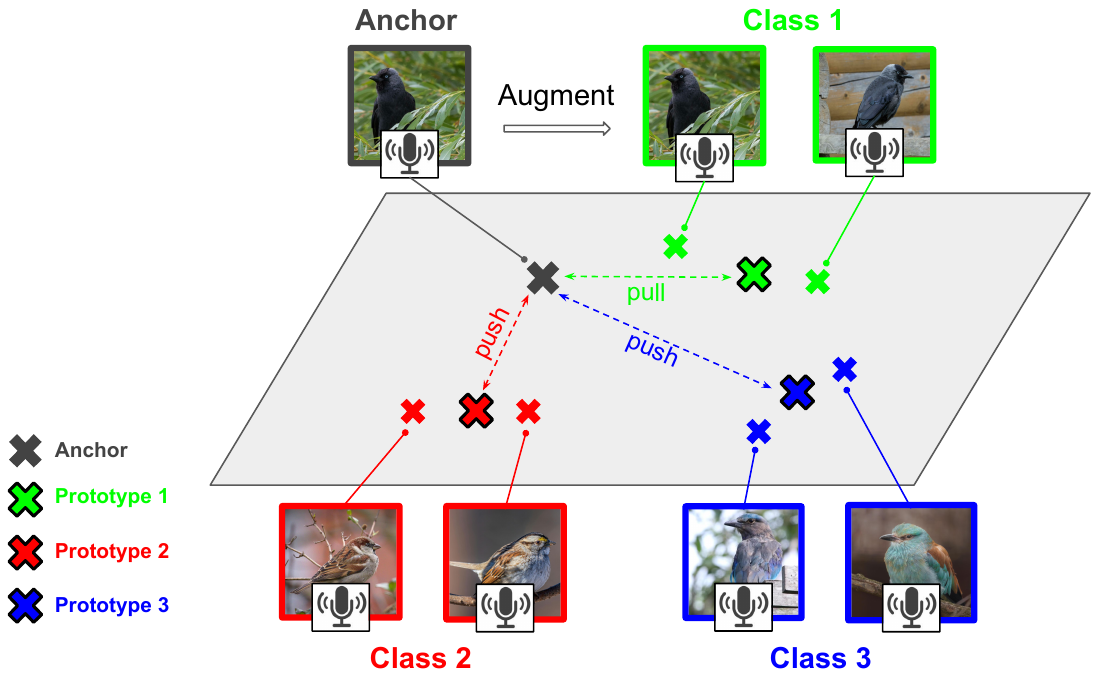}}
\caption{Prototypical contrastive pre-training of representations: Anchors are augmented and pulled towards the prototype of their class, while being pushed apart from prototypes of other classes.}
\label{fig:sslpip}
\end{figure}

\subsection{Supervised Contrastive Loss (SupCon)}

Given a batch with two views (transformations) of each example, let \( I \) denote the set of indices of all examples in the batch, \( P(i) \) represent the set of indices of positive examples for examples \( i \), and \( A(i) = I \setminus \{i\} \) represent the set of all other indices excluding \( i \). For an example \( i \), let \( {z}_i \) be its \( l_2 \)-normalized embedding and \( \tau \) the temperature parameter.

The SupCon~\cite{supcon} loss is defined as:
\begin{equation}
    \mathcal{L}^{\text{SupCon}} = \sum_{i \in I} \frac{-1}{|P(i)|} \sum_{p \in P(i)} \log \frac{\exp\left( {z}_i \cdot {z}_p / \tau \right)}{\sum_{a \in A(i)} \exp\left( {z}_i \cdot {z}_a / \tau \right)},
\end{equation}

The gradient of SupCon loss for an example \( i \) with respect to its embedding \( {z}_i \) is (please refer to \cite{supcon} for more details):
\begin{equation}
\label{eq:gradsupcon}
    \nabla_{{z}_i} \ell_{i}^{\text{SupCon}} = \frac{1}{|P(i)|} \sum_{p \in P(i)} \frac{1}{\tau} {z}_p - \frac{1}{\tau} \frac{\sum_{a \in A(i)} S_{ia} {z}_a}{\sum_{a \in A(i)} S_{ia}},
\end{equation}
where \( S_{ia} = \exp({z}_i \cdot {z}_a / \tau) \) is the similarity between \( {z}_i \) and \( {z}_a \).

This gradient consists of two terms: a positive term $\frac{1}{|P(i)|} \sum_{p \in P(i)} \frac{\displaystyle 1}{\displaystyle \tau} {z}_p$ pulling the embedding \( {z}_i \) towards its class centroid and a negative term $- \frac{\displaystyle 1}{\displaystyle \tau} \frac{\sum_{a \in A(i)} S_{ia} {z}_a}{ \sum_{a \in A(i)} S_{ia}}$ pushing it away from other examples.

\subsection{Prototypical Contrastive Loss (ProtoCLR)}

Motivated by the gradient of SupCon and drawing inspiration from prototypical learning~\cite{protonets, protonce}, we propose ProtoCLR (\textbf{Proto}typical \textbf{C}ontrastive \textbf{L}earning of \textbf{R}epresentations), which introduces class-level centroids into contrastive learning. The centroid for each class \( y \) in the batch is computed as
\(
    {c}_y = \frac{\displaystyle 1}{\displaystyle |C(y)|} \sum_{i \in C(y)} {z}_i
\), where \( C(y) \) is the set of indices of examples with label \( y \) and \( |C(y)| \) is its size. We define the ProtoCLR loss as:
\begin{equation}
    \mathcal{L}^{\text{ProtoCLR}} = \sum_{i \in I} \log \frac{\exp\left( {z}_i \cdot {c}_{y_i} / \tau \right)}{\sum_{y \in Y} \exp\left( {z}_i \cdot {c}_y / \tau \right)},
\end{equation}

\noindent where \( {c}_{y_i} \) is the centroid of the class to which example \( i \) belongs, and \( Y \) is the set of all classes in the batch.

Similarly to SupCon, the gradient for ProtoCLR is:

\begin{equation}
    \nabla_{{z}_i} \mathcal{L}^{\text{ProtoCLR}}_{i} = \frac{1}{\tau}  {c}_{y_i} - \frac{1}{\tau} \frac{\sum_{y \in Y} S_{iy} {c}_y}{\sum_{y \in Y} S_{iy}} ,
\end{equation}
where \( S_{iy} = \exp({z}_i \cdot {c}_y / \tau) \) is the similarity between \( {z}_i \) and \( {c}_y \).

The positive term remains the same as in the gradient of SupCon, pulling the embeddings \( {z}_i \) towards the centroids of their respective classes. The difference is the negative term $\frac{\displaystyle 1}{\displaystyle \tau} \frac{\sum_{y \in Y} S_{iy} {c}_y}{\sum_{y \in Y} S_{iy}}$: in ProtoCLR, the embeddings are pushed away from the weighted average of the centroids as opposed to the individual embeddings in SupCon.

\subsection{ProtoCLR vs SupCon}

In order to comprehensively assess the efficacy of ProtoCLR, we conduct a comparative analysis with SupCon.

\subsubsection{Complexity}

SupCon has a computational cost of \( \mathcal{O}(N^2) \) due to computing dot products between all pairs of examples in a batch of size \( N \), independent of the number of classes \( C \). In contrast, ProtoCLR reduces this to \( \mathcal{O}(N \times C) \) by computing dot products with class prototypes. Since \( C \) is smaller or equal to \( N \), ProtoCLR is more efficient.

\subsubsection{Variance}

SupCon relies on pairwise comparisons within the same class, which can lead to higher variance due to intra-class variability. In contrast, ProtoCLR compares embeddings \( {z}_i \) with class prototypes \( {c}_{y_i} \), reducing variance as the prototype variance \( \text{Var}({c}_{y_i}) \) decreases by \( N_{y_i}^2 \), where \( N_{y_i} \) is the number of examples in class \( y_i \). This leads to lower noise and more stable gradients in ProtoCLR:

\[
\text{Var}({c}_{y_i}) = \frac{\text{Var}(\sum_{j \in y_i} {z}_j)}{N_{y_i}^2}.
\]

\subsubsection{Near Convergence Equivalence}

Near convergence, both SupCon and ProtoCLR promote intra-class compactness with embeddings clustering tightly around class centroids:
\(
  {z}_i \approx {c}_{y_i} \quad \text{for all } i \in I
\).
In SupCon, the negative term in eq.~(\ref{eq:gradsupcon}) can be rewritten as:
\begin{equation}
  \frac{S_{ia} {z}_a}{S_{ia}} \approx \frac{\exp({z}_i \cdot {z}_{a} / \tau) {z}_{a}}{\exp({z}_i \cdot {z}_{a} / \tau)} \approx \frac{\exp({z}_i \cdot {c}_{y_a} / \tau) {c}_{y_a}}{\exp({z}_i \cdot {c}_{y_a} / \tau)} \approx \frac{S_{iy_a} {c}_{y_a}}{S_{iy_a}}.
\end{equation}

\noindent Thus, SupCon and ProtoCLR converge to similar strategies, ensuring intra-class compactness and inter-class separability in the final learned representations.

\section{Experiments}

\begin{table}[!h]
\centering
\begin{tabular}{llccc}
\midrule
Split & Dataset & \# of recordings  & \# of classes \\
\midrule
\multirow{1}{*}{Train} & XC & 684,744 & 10,127 \\
\midrule
\multirow{1}{*}{Val} & POW & 16,052 & 48 \\
\midrule
\multirow{7}{*}{Test} & PER & 14,798 & 132 \\
& NES & 6,952 & 89 \\
& UHH & 59,583 & 27 \\
& HSN & 10,296 & 19 \\
& SSW & 50,760 & 96 \\
& SNE & 20,147 & 56 \\
\midrule
\end{tabular}
\caption{Overview of the datasets in the BIRB benchmark.}
\label{tab:benchmark}
\end{table}

\begin{table*}[!htbp]
\centering
\begin{tabular}{lcccccccc}
\midrule
\textbf{Model} & \textbf{Training Params (M)} & \textbf{PER} & \textbf{NES} & \textbf{UHH} & \textbf{HSN} & \textbf{SSW} & \textbf{SNE} & Mean \\
\midrule
\textbf{\deemph{Random Guessing}} & - & \deemph{0.75} & \deemph{1.12} & \deemph{3.70} & \deemph{5.26} & \deemph{1.04} & \deemph{1.78} & \deemph{2.22} \\
\midrule
\textbf{One-Shot Classification} & & & & & & & &  \\
\midrule
BirdAVES-biox-base & 95 & 7.41$\pm$1.0 & 26.4$\pm$2.3 & 13.2$\pm$3.1 & 9.84$\pm$3.5 & 8.74$\pm$0.6 & 14.1$\pm$3.1 & 13.2 \\
BirdAVES-bioxn-large & 316 & 7.59$\pm$0.8 & 27.2$\pm$3.6 & 13.7$\pm$2.9 & 12.5$\pm$3.6 & 10.0$\pm$1.4 & 14.5$\pm$3.2 & 14.2 \\
BioLingual & 153 & 6.21$\pm$1.1 & 37.5$\pm$2.9 & 17.8$\pm$3.5 & 17.6$\pm$5.1 & 22.5$\pm$4.0 & 26.4$\pm$3.4 & 21.3 \\
Perch & 80 & 9.10$\pm$5.3 & 42.4$\pm$4.9 & 19.8$\pm$5.0 & 26.7$\pm$9.8 & 22.3$\pm$3.3 & 29.1$\pm$5.9 & 24.9 \\
\cmidrule(lr){1-1}
Ours & & & & & & &  \\
\cmidrule(lr){1-1}
CE & 23 & 9.55$\pm$1.5 & 41.3$\pm$3.6 & 19.7$\pm$4.7 & 25.2$\pm$5.7 & 17.8$\pm$1.4 & 31.5$\pm$5.4 & 24.2 \\
SimCLR & 19 & 7.85$\pm$1.1 & 31.2$\pm$2.4 & 14.9$\pm$2.9 & 19.0$\pm$3.8 & 10.6$\pm$1.1 & 24.0$\pm$4.1 & 17.9 \\
SupCon & 19 & 8.53$\pm$1.1 & 39.8$\pm$6.0 & 18.8$\pm$3.0 & 20.4$\pm$6.9 & 12.6$\pm$1.6 & 23.2$\pm$3.1 & 20.5 \\
ProtoCLR & 19 & 9.23$\pm$1.6 & 38.6$\pm$5.1 & 18.4$\pm$2.3 & 21.2$\pm$7.3 & 15.5$\pm$2.3 & 25.8$\pm$5.2 & 21.4 \\
\midrule
\textbf{Five-Shot Classification} & & & & & & & & \\
\midrule
BirdAVES-biox-base & 95 & 11.6$\pm$0.8 & 39.7$\pm$1.8 & 22.5$\pm$2.4 & 22.1$\pm$3.3 & 16.1$\pm$1.7 & 28.3$\pm$2.3 & 23.3 \\
BirdAVES-bioxn-large & 316 & 15.0$\pm$0.9 & 42.6$\pm$2.7 & 23.7$\pm$3.8 & 28.4$\pm$2.4 & 18.3$\pm$2.3 & 27.3$\pm$2.3 & 25.8 \\
BioLingual & 153 & 13.6$\pm$1.3 & 65.2$\pm$1.4 & 31.0$\pm$2.9 & 34.3$\pm$3.5 & 43.9$\pm$0.9 & 49.9$\pm$2.3 & 39.6 \\
Perch & 80 & 21.2$\pm$1.2 & 71.7$\pm$1.5 & 39.5$\pm$3.0 & 52.5$\pm$5.9 & 48.0$\pm$1.9 & 59.7$\pm$1.8 & 48.7 \\
\cmidrule(lr){1-1}
Ours & & & & & & & & \\
\cmidrule(lr){1-1}
CE & 23 & 21.4$\pm$1.3 & 69.2$\pm$1.8 & 35.6$\pm$3.4 & 48.2$\pm$5.5 & 39.9$\pm$1.1 & 57.5$\pm$2.3 & 45.3 \\
SimCLR & 19 & 15.4$\pm$1.0 & 54.0$\pm$1.8 & 23.0$\pm$2.3 & 32.8$\pm$4.0 & 22.0$\pm$1.2 & 40.7$\pm$2.4 & 31.3 \\
SupCon & 19 & 17.2$\pm$1.3 & 64.6$\pm$2.4 & 34.1$\pm$2.9 & 42.5$\pm$2.9 & 30.8$\pm$0.8 & 48.1$\pm$2.4 & 39.5 \\
ProtoCLR & 19 & 19.2$\pm$1.1 & 67.9$\pm$2.8 & 36.1$\pm$4.3 & 48.0$\pm$4.3 & 34.6$\pm$2.3 & 48.6$\pm$2.8 & 42.4 \\
\midrule
\end{tabular}
\caption{Top-1 accuracy for one-shot and five-shot classification. All reported results are the average of ten runs.} 
\label{tab:results}
\end{table*}

\subsection{Few-Shot Classification Benchmark}

The original task of the benchmark BIRB~\cite{birb} focuses on information retrieval in bioacoustics, aiming to retrieve bird vocalizations from passively recorded datasets using focal recordings for training. Downstream datasets may contain long audio recordings where events of interest are sparse. Individual bird events are detected using peak detection~\cite{denton}, which identifies frames with high energy to used for the task. We build on previous work~\cite{moummadicasspw}, which shows that using a pretrained classifier on AudioSet~\cite{audioset} is a good proxy for selecting windows with the highest bird class activation.

Following the MetaAudio framework, we focus on a multi-class classification task to define a few-shot evaluation for assessing the generalization capabilities of pre-trained models~\cite{buroojtransfer}. BIRB provides a large training set, XC (Xeno-Canto), consisting of focal recordings from nearly 10,000 species, while the validation and test datasets contain soundscape recordings. The benchmark details are provided in Table~\ref{tab:benchmark}.

Despite the inherently multi-label nature of soundscape recordings, we adopt a mono-class assumption for simplicity and consistency with the methodology used in BIRB~\cite{birb}. To mitigate the effects of label confusion, we sample few-shot examples 10 times using different seeds. 

Following BioCLIP~\cite{bioclip}, we sample \textit{k}-shot learning tasks by randomly selecting \textit{k} examples for each class and obtain the audio embeddings from the audio encoder of the pre-trained models. We then compute the average feature vector of the \textit{k} embeddings as the training prototype for each class. All the examples left in the dataset are used for testing.

To make predictions, we employ SimpleShot \cite{SimpleShot} by applying mean subtraction and L2-normalization to both centroids and test feature vectors. We then select the class whose centroid is closest to the test vector as the prediction. We repeat each few-shot experiment 10 times with different random seeds and report the mean and standard deviation accuracy in Table~\ref{tab:results}.

\subsection{Reference Systems}

To compare supervised contrastive approaches SupCon and ProtoCLR for addressing domain generalization, we train a reference system using cross-entropy (CE) loss as a supervised baseline and SimCLR as a self-supervised contrastive baseline. Additionally, we evaluate large-scale, state-of-the-art models in bioacoustics: BirdAVES~\cite{aves}, a transformer model based on the speech model HuBERT, trained in a self-supervised manner based on the AVES framework~\cite{aves}. We evaluate two versions of BirdAVES, biox-base, a 12-layer transformer with 768 units, trained on Xeno-Canto~\cite{xc} and animal sounds from AudioSet~\cite{audioset} and VGGSound~\cite{vggsound}, and bioxn-large, a 24-layer transformer with 1024 units, trained on the same datasets plus iNaturalist~\footnote{\url{https://github.com/gvanhorn38/iNatSounds}}. BirdAVES outputs one embedding vector per time frame, which we average across the time dimension to produce a single vector representing the entire audio clip; the encoder of BioLingual~\cite{biolingual}, an audio transformer pre-trained on AudioSet and fine-tuned using contrastive language-audio training to align animal sounds with text captions describing the class across a large collection of data including Xeno-Canto, iNaturalist, Animal Sound Archive, \dots etc; and the encoder of Perch~\cite{birb}, a large EfficientNet~\cite{efficientnet} trained on Xeno-Canto for species classification, as well as taxonomic ranks genus, family, and order. We evaluate only models trained specifically on bird sounds, as general-purpose audio models have been shown to perform poorly in this domain~\cite{buroojtransfer}. Similarly, Hamer et al.~\cite{birb} found that general-purpose models can, in some cases, perform worse than simply averaging mel spectrogram features without any learning.

\subsection{Pre-training Details}

We follow the same preprocessing as Moummad et al.~\cite{moummadicasspw}. We train all models with CvT-13~\cite{cvt}, an efficient 2D transformer architecture with 20M parameters, on XC for 300 epochs using the AdamW optimizer with a batch size of 256, with a weight decay of $1 \times 10^{-6}$. Following Moummad et al.~\cite{moummadicasspw}, we apply the domain-agnostic data augmentations found to be effective for bird sound representations: circular time shift~\cite{uclser}, SpecAugment~\cite{specaug}, and spectrogram mixing~\cite{byola}.
These models are trained with a projector of dimension 128. For the CE loss, we only apply circular time shift and SpecAugment as augmentations, excluding Spectrogram Mixing, as it prevented the model from converging. The learning rate for CE and ProtoCLR is set to $5 \times 10^{-4}$, while for SupCon and SimCLR, we use a learning rate of $1 \times 10^{-4}$. We tune hyperparameters by monitoring \textit{k}-NN accuracy on the POW validation dataset, which is split randomly into a training and a validation subset.

\subsection{Results and Discussion}

Table~\ref{tab:results} presents the performance of different models on 1-shot and 5-shot bird sound classification tasks. ProtoCLR outperforms SupCon, while being computationally efficient during training: 1 epoch on XC with a batch size of 256, SupCon requires 80.4B multiply-accumulate operations, whereas ProtoCLR requires only 28.3B. 

For self-supervised methods, SimCLR outperforms BirdAVES-biox-base and BirdAVES-bioxn-large in both 1-shot and 5-shot learning, despite using a smaller model and simple data augmentation. This suggests that invariant learning is more effective than self-prediction methods for few-shot scenarios, consistent with findings in computer vision~\cite{msn}.

On average, ProtoCLR outperforms BioLingual in both 1-shot learning and 5-shot learning. However, when examining individual datasets, BioLingual performs worse on PER, NES, UHH, and HSN, yet significantly better on SSW and SNE. Interestingly, certain models appear to specialize in specific species, suggesting a promising research direction: distilling these specialized models into a single general model that can leverage the strengths of each.

Perch outperforms all models in both 1-shot and 5-shot classification across all datasets, except for PER, where it performs slightly worse than CE. This suggests that incorporating taxonomic ranks as auxiliary tasks can enhance the discriminative capabilities of the feature extractor.

\section{Conclusion}

In this work, we addressed the challenge of domain generalization for bird sound classification in few-shot scenarios, with a focus on the domain shift from focal to soundscape recordings. We proposed a new few-shot evaluation, derived from the BIRB datasets, to evaluate the generalization capabilities of models trained on focal recordings and tested on soundscapes. Additionally, we introduced ProtoCLR, a computationally efficient alternative to SupCon, inspired by prototypical learning. ProtoCLR maintains performance levels comparable to SupCon, and in some cases, outperforms it.

SimCLR is shown to outperform state-of-the-art self-supervised models. It falls behind supervised models but offer a scaling advantage when large unlabeled data is available. Without strong data augmentation, CE outperforms supervised contrastive methods. Future work will explore data augmentation for bioacoustics to efficiently leverage invariant learning. Incorporating taxonomic ranks into the learning process holds promise as shown by Perch. One potential limitation of ProtoCLR, as well as other state-of-the-art approaches, is the assumption that each example belongs to a single class. This is often not the case in bioacoustics, where sounds can overlap in soundscape recordings. Future work should consider an extension of ProtoCLR that allows examples to belong to multiple classes.





\bibliographystyle{IEEEbib}
\bibliography{refs}


\end{document}